\documentclass[twocolumn]{article}
\usepackage{babel}
\usepackage{multirow}
\usepackage{listings}
\usepackage{booktabs}
\usepackage{graphicx}
\usepackage{adjustbox}
\title{Time-to-Lie: Identifying Industrial Control System Honeypots Using the Internet Control Message Protocol}

\author{
Jacob Williams\\
\textit{\small jacobjohn.williams@bristol.ac.uk}\\
\and
Matthew Edwards\\
\textit{\small matthew.john.edwards@bristol.ac.uk}\\
\and
Joseph Gardiner\\
\textit{\small joe.gardiner@bristol.ac.uk}\\
}

\begin{document}
\date{}
\maketitle
\begin{abstract}
The convergence of information and operational technology networks has created previously unforeseen security issues. To address these issues, both researchers and practitioners have integrated threat intelligence methods into the security operations of converged networks, with some of the most valuable tools being honeypots that imitate industrial control systems (ICS). However, the development and deployment of such honeypots is a process rich with pitfalls, which can lead to undiagnosed weaknesses in the threat intelligence being gathered.
This paper presents a side-channel method of covertly identifying ICS honeypots using the time-to-live (TTL) values of target devices. We show that many ICS honeypots can be readily identified, via minimal interactions, using only basic networking tools. In a study of over 8,000 devices presenting as ICS systems, we detail how our method compares to an existing honeypot detection approach, and outline what our methodology reveals about the current population of live ICS honeypots. In demonstrating our method, this study aims to raise awareness of the viability of the TTL heuristic and the prevalence of its misconfiguration despite its presence in literature.     
\end{abstract}

\section{Introduction} 
The industrial control system (ICS) is a broad category of devices used to provide control and monitoring functionalities to manufacturing and industrial facilities~\cite{KNAPP20159}. These deployments consist of typical networking and computer systems alongside unique and process specific hardware like sensors, actuators, and logical controllers which are robust, time sensitive devices meant for precise process control. These processes often form what is known as `Critical Infrastructure' and can oversee systems that are important for the day-to-day functioning of society. This includes processes like food production, electrical generation and supply, and water purification. In the past these systems were secured through their obscurity, they were often kept separate from each other and were not connected to the internet. The proliferation of wide-area networks, rise of inter-connectivity, and the commodification of data in recent years has caused a convergence of these previously separate systems. As a result, there are devices with poor security features being exposed to the internet or to vectors of attack that were never considered in their design. 

There exists a two-fold problem when securing ICS. The first is a significant lack of workable threat intelligence and the second is a gap in deployable network security systems. An attempt to address both of these problems concerns the creation of ICS honeypots, deceptive pseudo-devices that masquerade as functioning pieces of equipment to fool threat actors into attacking them while recording their every move. These come in two forms, the research honeypot addresses the former problem by being exposed to the internet to capture interactions with opportunistic attackers and produce threat intelligence. The security honeypot works at the latter issue by being placed within a network as a trip-wire meant to delay, confuse, or even repel attackers. 

Honeypot are already commonly used in fields such as the internet-of-things, to track and counteract botnet activity. ICS security is a budding field and considerably understudied, with ICS honeypots being even more neglected. Recent calls to action have seen a rapid increase in ICS centered research following a number of high-profile attacks against critical infrastructure such as Stuxnet~\cite{Stuxnet_dossier}, CrashOverride~\cite{crashoverride, geiger2020analysis}, and BlackEnergy3~\cite{geiger2020analysis}. One strand of this research activity focuses on the creation of ICS honeypot implementations that are capable of deceiving even the most sophisticated actors. These efforts often take on two forms: the development of additional honeypot features, and the exploration of identification techniques that can inform development. This study fits into the latter pattern, highlighting a major weakness of many current ICS honeypot implementations. Studying basic device properties reveals that many honeypot systems fail to imitate the standard packet time-to-live (TTL) values reported by real ICS device despite its coverage in literature~\cite{Buza_2014_CryPLH_Smart_Energy, Haney_2014_SCADA_honeynet, Holczer_2015_plc_honeypot, Mashima_2020_smart_grid_practical_threat_intel, Feng_2022, Pace_2023}. Leveraging this to advance the development of industrial honeypots, this study contributes: \begin{itemize}
    \item Demonstration of the use of TTL values as a honeypot identification heuristic.
    \item Large-scale evaluation against real world devices discovered leveraging Shodan.
    \item A comparative analysis using the Shodan honeyscore system to evaluate the performance of our implementation and the honeyscore system itself.
\end{itemize}

The remainder of this paper proceeds as follows. Section~\ref{sec:lit} explores previous industrial honeypot literature, detailing the various detection methods and the solutions authors have supplied. Following this, Section~\ref{sec:method} details the approach and tools used for sampling purported ICS devices, the specifics of ping and traceroute testing, reconstructing a target's time-to-live value, and the processing of results. Section~\ref{sec:results} performs a comparative analysis using the retrieved time-to-live values and the records pulled from Shodan to identify areas of consensus and contention. This is then concluded in Section~\ref{sec:discuss} and \ref{sec:conclusion} discussing the meaning of the results, a review of the methods used, and the identification of further work.

\section{Related Work} \label{sec:lit}
The industrial honeypot literature concerns the creation of honeypots deceptive enough to fool threat actors at differing levels of sophistication. The literature tends to study either the shortcomings of established honeypots or the development of more effective solutions, a body of work which can itself be further divided according to whether the honeypot is intended as an exposed research honeypot or a network-based security solution. Only a handful of studies acknowledge the potential exposure of honeypots via TTL values~\cite{Buza_2014_CryPLH_Smart_Energy, Haney_2014_SCADA_honeynet, Holczer_2015_plc_honeypot, Mashima_2020_smart_grid_practical_threat_intel, Feng_2022, Pace_2023} and proceed to implement changes.

The easiest method to reveal honeypots is finding static characteristics~\cite{Zamiri_2019_gas_what, Kuman_2017_IMUNES_CONPOT, Ferretti_2019_background_noise} or deterministic values~\cite{Zamiri_2019_gas_what, kendrick_2019_energy, Lopez-Morales_2020_HoneyPLC}. Static characteristics are revealed through scanning and include system names, locations, and module numbers that are typical of default ConPot configurations~\cite{Zamiri_2019_gas_what}. Honeypots that implement changes to these characteristics remain `low interaction' but are less easily discovered~\cite{Kuman_2017_IMUNES_CONPOT, Ferretti_2019_background_noise}. Deterministic values originate from the simulation of processes or lack thereof. Deterministic value fluctuations are unlikely in deployments~\cite{Zamiri_2019_gas_what} and the simulations in honeypots need to be fine tuned to avoid this flaw, which involves creating sophisticated algorithms and tools to accurately simulate complex processes~\cite{Lopez-Morales_2020_HoneyPLC, Kuman_2017_IMUNES_CONPOT, Redwood_2015_symbolic_honeynet_framework, Antonioli_2016_honeypots-in-a-box, Xiao_2018_S7CommTrace, Mashima_2020_smart_grid_practical_threat_intel, kendrick_2019_energy}. 


Achieving higher levels of deception and interactivity requires creating sophisticated protocol simulations. Honeypots are commonly missing complex protocol functions~\cite{Zamiri_2019_gas_what, Grigoriou_protecting_IEC60870-5-104}, producing irregular responses~\cite{Zamiri_2019_gas_what, Lopez-Morales_2020_HoneyPLC, Vetterl_2018_Bitter_Harvest, Xiao_2018_S7CommTrace}, and exhibiting inconsistent scanning results~\cite{Mashima_2020_smart_grid_practical_threat_intel}. Simply producing better implementations is problematic due to the propriety nature of industrial protocols like S7CommTrace, which is primarily understood through reverse engineering efforts. Non-industrial protocols like SSH, TelNet, and HTTP(S) are still capable of revealing honeypots~\cite{Zamiri_2019_gas_what} if the basic implementation included with tools like ConPot is retained. Better interactivity and deception is produced by combining multiple honeypot technologies in order to account for their different weaknesses~\cite{Mashima_2020_smart_grid_practical_threat_intel, Rashid_2020_Know_your_enemy, Hilt_2020_caught_in_the_act, Redwood_2015_symbolic_honeynet_framework} but this can prove detrimental if too many are used on the same host~\cite{Rashid_2020_Know_your_enemy}. 

Side-channel honeypot detection methods involve interrogating the honeypot to find artefacts of the underlying platform or operating system that leak through. The most common in this category involves exposing cloud hosts by analysing domains and IP address ranges~\cite{Zamiri_2019_gas_what}. A common work-around to this problem is to use a lesser known hosting provider~\cite{Xiao_2018_S7CommTrace,Khoury_2020_IoT_malware_at_scale}, but how effective this has been is hard to determine. A similar problem arises from university hosted honeypots~\cite{kendrick_2019_energy, Buza_2014_CryPLH_Smart_Energy, Ferretti_2019_background_noise}. Sophisticated threat actors will conduct open source intelligence gathering on domains or addresses before beginning any active reconnaissance to ensure the validity of the target. SSH is a common offender revealing operating systems through client strings~\cite{Vetterl_2018_Bitter_Harvest} and exchanged banners~\cite{Jicha_2016_analysis_of_conpot} in systems that have been configured remotely. Similarly, creating a honeyweb with a centralised log store exposes ports and services not found in industrial systems, such as port 514 for SysLog~\cite{Jicha_2016_analysis_of_conpot}, port 5000 for LogStash, or Kibana for visualisation on port 5601~\cite{Navarro_2019_gathering_intel}. 

\section{Method} \label{sec:method}

The TTL field of IPv4 packets is a counter for limiting packet lifetimes. It is a static value determined by operating systems that is decremented every time the packet is routed~\cite{Tanenbaum_computer_networks}. When this value reaches zero the packet is discarded and a warning is sent to the source host. The default maximum TTL value can be used to identify the operating system of a device, since the values differ across deployments. For example, the default TTL for Linux and Windows is 64 and 128 respectively~\cite{OSTechnix_TTL}. The same devices across multiple generations can also have differing TTLs. This is demonstrable by Siemens S7 devices as shown in Table~\ref{tab:device_ttl_values}. 

The scope of this study is limited to Siemens devices due to their centrality to the honeypot literature, as explored in Section~\ref{sec:lit}. Siemens is a market leader in industrial manufacturing~\cite{Reuters_siemens_growth}, in the top five for market shares globally~\cite{MordorInt_Siemens_market_leader}, and is the largest producer of industrial equipment in Europe~\cite{CoherentMarketInsights_Siemens_Europe}. This study began by querying various Siemens devices in the S7 range for their default TTL value, recorded alongside the device models in Table~\ref{tab:device_ttl_values}, models are variations of the same device with different configurations. Note that most default TTL values differ significantly from the Linux and Windows defaults. Comparing the TTL values of a device as elicited from a remote probe to the known values in Table~\ref{tab:device_ttl_values} provides a signal regarding whether or not a machine claiming to be Siemens S7 device is likely to be a honeypot.

\begin{table}
    \centering
    \begin{tabular}{c c c}
         Range & Model & TTL \\
         \hline
         ET200S & 6ES7 151-8AB00-0AB0 & 30 \\
         \midrule 
         S7-300 & 6ES7 322-1BH01-0AA0 & 60 \\
         \midrule 
         \multirow{3}{3.2em}{S7-1200} & 6ES7 212-1BE40-0XB0 & 30 \\
          & 6ES7 214-1AG40-0XB0  & 30 \\
          & 6ES7 215-1AG40-0XB0  & 30 \\
          \midrule
         S7-1500 & 6ES7 522-1BL10-0AA0 & 255 \\
         \bottomrule
    \end{tabular}
    \caption{Model and TTL value of reference Siemens devices.}
    \label{tab:device_ttl_values}
\end{table}

\subsection{Gathering Devices} \label{subsec:gathering_devices}
The search engine Shodan was used to gather the profiles of purported live and accessible S7 devices that could be queried for their TTL value. Shodan regularly scans the internet for devices with ports exposed to the internet, these are recorded in a publicly searchable database. Shodan allows for searching with strings that correlate with metadata in ports recorded, this enables the gathering of Siemens devices using the `6ES7' string, which denotes the series of S7 devices including the S7-200, 300, 400, 1200, and 1500. 

This dataset was expanded by including known honeypot strings identified in previous research~\cite{Zamiri_2019_gas_what, Kuman_2017_IMUNES_CONPOT}. The intent of this is to include known honeypots in the dataset to help evaluate the effectiveness of the TTL value as a heuristic for honeypot identification. By first applying our TTL testing methodology, and then querying the Shodan records for presence of the `honeypot' tag that is applied by the Honeyscore system, we allow for a comparison of results between the two heuristics. The common honeypot strings utilised were 'Technodrome', 'Mouser Factory', and '[00:13:EA:00:00:00]', all of which occur in a variety of ICS protocol simulations.


Strings are searched sequentially and results were stored in a JSON format for easier processing. Each device was tagged with the Shodan query that first yielded it, for traceability purposes. Shodan searches yield comprehensive results that include addresses, the port that contained matching metadata, and tags that Shodan assigns to devices using their own algorithms. This data was stored to allow additional analyses of devices to be performed later. After processing and analysis, the address section of the dataset was anonymised with a SHA-256 hashing function. 

\subsection{Time-to-Live Testing}
Once compiled, the dataset was ingested by an additional Python script which performed the Ping and TraceRoute operations on each address. These were recorded individually, then used to reassemble an estimated maximum TTL. Ping yields the TTL of the target device after it has traversed back to the source host and TraceRoute counts the steps between the source and destination hosts. Combining these results produces the reconstructed TTL value, which can then be compared against the expected values in Table~\ref{tab:device_ttl_values} and the standard operating system values~\cite{OSTechnix_TTL}. Ping and TraceRoute use either Universal Datagram Protocol (UDP) or the Internet Control Messaging Protocol (ICMP) and are therefore minimally invasive. This creates a covert method of identification that does not impact the run-time integrity of the device. The asymmetric nature of internet routing means that reconstruction may not always be exact, therefore the reconstructed value is compared to its nearest TTL value entry. If the nearest value matches a real device, the device is marked as not being a honeypot per the TTL heuristic, which we label `Local'. If the nearest match is an operating system, the heuristic flags this device as a potential honeypot. 
The results of this process are compared against the Shodan Honeyscore system. Two situations arise in the following analysis, where the Local and Shodan results either achieve consensus about whether a device is a honeypot or are found to be in contention. Further analysis uses the Shodan records to describe the characteristics of devices that most commonly appear in either category, and discusses other evidence that allows for contentious cases to be resolved.

An Amazon AWS EC2 t2.micro instance possessing 1 vCPU, 1GB RAM, and 10GB of solid state memory was utilised for this testing. The scripts were run atop a headless 64-bit Debian 12 operating system. The instance made use of a single network interface configured to allow all ICMP traffic to and from it, as well as key authenticated SSH access. This interface was associated with a unique elastic IP in the same region as the virtual machine, in EU North.

\section{Results} \label{sec:results}

\begin{figure}
    \centering
    \includegraphics[scale=0.37]{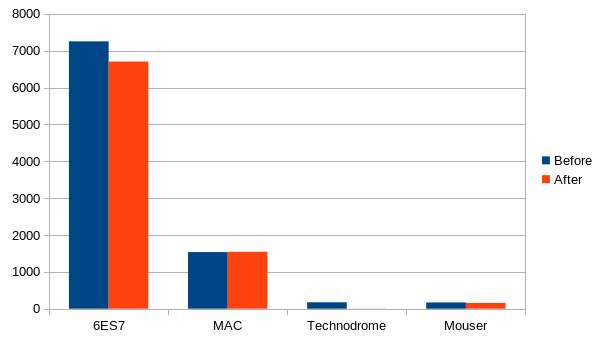}
    \caption{Total search results before and after duplicate reduction.}
    \label{fig:dupe_reduction}
\end{figure}

Four Shodan search queries generated a total set of 9132 devices. `6ES7' yielded the most results with 7252 entries, as viewable in Figure~\ref{fig:dupe_reduction} it provided the most addresses by a significant margin. The MAC address search string was the second most effective and created 1536 addresses. This was then followed by `Technodrome' and `Mouser' which produced 174 and 170 results respectively. In order to reduce the amount of tests necessary the dataset was pre-processed to remove duplicate entries. This created a reduction of 722 entries as shown in Figure~\ref{fig:dupe_reduction}, leaving a dataset containing 8410 unique devices. 

\begin{figure}
    \centering
    \includegraphics[scale=0.3]{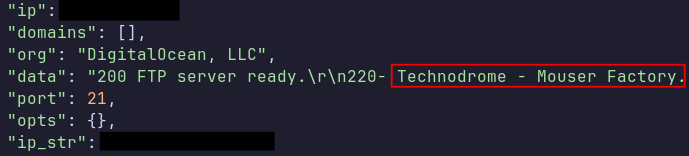}
    \caption{Technodrome and Mouser strings appearing in the same result.}
    \label{fig:technomouser}
\end{figure}

The `6ES7' string saw a considerable reduction in results, but the most obviously reduced is `Technodrome'. The significant reduction it received is most likely due to its characteristic overlap with `Mouser' given both are typically present in honeypot FTP strings as seen in Figure~\ref{fig:technomouser}. Inversely, MAC saw no considerable reduction due to only being present in telnet honeypots.  

These strings were processed by a script performing the Ping and TraceRoute tests. This caused a reduction of 3269 entries due to errors, leaving a total of 5141 valid results. Errors occurred where either the ping or traceroute functions fail to receive a result. The script performs both UDP and ICMP tests, if both receive no reply then the device is truly unreachable. If one test fails it is impossible to reassemble the final TTL of the address and the entry is discarded from analysis. Why an address is not reachable is debatable, given that it is recorded on Shodan it can be assumed it was reachable at some point. This would mean that the device has either been taken offline or has been configured to not respond to UDP or ICMP packets. 

\begin{figure}
    \centering
    \includegraphics[scale=0.39]{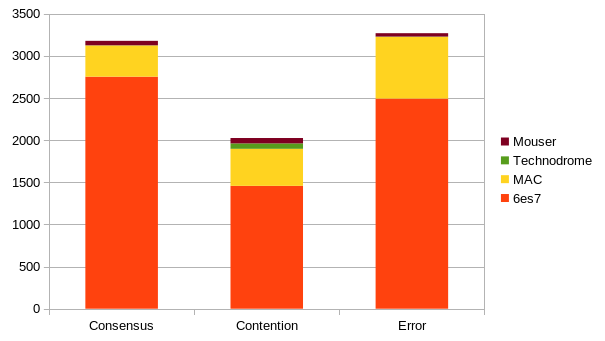}
    \caption{Consensus, Contention, and Error results visualised by distribution between search strings.}
    \label{fig:results-by-string}
\end{figure}

Figure~\ref{fig:results-by-string} shows no singular search string occupying a share of errors disproportional to its total results. Figure~\ref{fig:results-by-string} demonstrates potential use of the TTL heuristic, with the amount the two algorithms reach consensus being 1218 higher than when they are in contention. There were a total of 3176 devices where both heuristics agreed upon a classification, with contention seen for 1958 devices. 

\subsection{Consensus Results}

\begin{figure}
    \centering
    \includegraphics[scale=0.37]{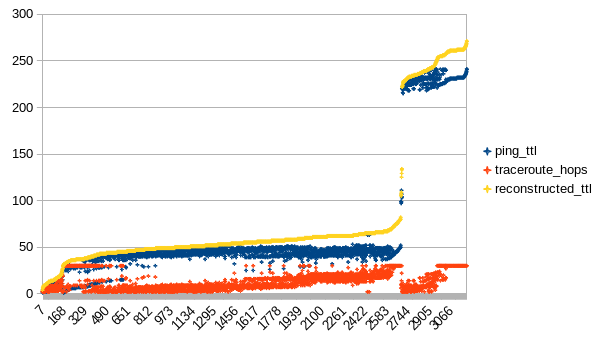}
    \caption{Consensus ping, traceroute, and reconstructed TTL results in linear ascending order across 3176 results.}
    \label{fig:consensus_overview}
\end{figure}

Figure~\ref{fig:consensus_overview} displays the variety of observed values for the consensus results. The consensus results primarily occupy the ranges of 0 to 60 and 200 to 250 where both the Local and Shodan algorithms have identified them as being legitimate S7 devices. Of our 3176 consensus results, we see 2811 addresses deemed to host legitimate devices.  We then see a notable jump between 60 to 200 where results for Linux and Windows hosts fall, showing the difference in quantity between honeypots and legitimate devices. 

Figure~\ref{fig:consensus_overview} communicates a parallel progression of Ping TTL with the final reconstructed TTL, but the TraceRoute TTL remaining stable. The Ping value will always be closer to the original TTL value and the TraceRoute will always be the lesser of the two results. The TraceRoute results show a number of obvious plateaus attributable to the default TraceRoute TTL maximum being 30 hops before returning a value. 

Analysing Shodan records reveals the numerous reasons the honeyscore system concluded the devices are honeypots. The port distribution of the results in Table~\ref{tab:consensus_port_results} shows which features revealed the honeypot. Port 21 and 2121 fall under the same FTP bracket, with the former being the typically used port and latter being a proxy port. Both of these ports reveal the honeypot via common honeypot strings that begin with `200 FTP server ready. 220- Technodrome - Mouser Factory.' which is present in ConPot and its forks. Port 50100 presents another common string with `Welcome...Connected to [00:13:EA:00:00:00]' attributed to the ConPot Telnet implementation. 

\begin{table}
    \centering
    \begin{tabular}{ c c c }
         Port & No. of Occurrences & \% of Consensus \\
         \midrule
         21 & 52 & 1.64\% \\
         102 & 2400 & 75.57\% \\
         161 & 344 & 10.83\% \\
         587 & 1 & 0.03\% \\
         1701 & 1 & 0.03\% \\
         2121 & 1 & 0.03\% \\
         4840 & 9 & 0.28\% \\
         50100 & 360 & 11.33\% \\
         \bottomrule
    \end{tabular}
    \caption{Distribution of consensus port results based on the number of occurrences and the percentage of results they occupy.}
    \label{tab:consensus_port_results}
\end{table}

\begin{figure}[ht]
    \centering
    \includegraphics[scale=0.2]{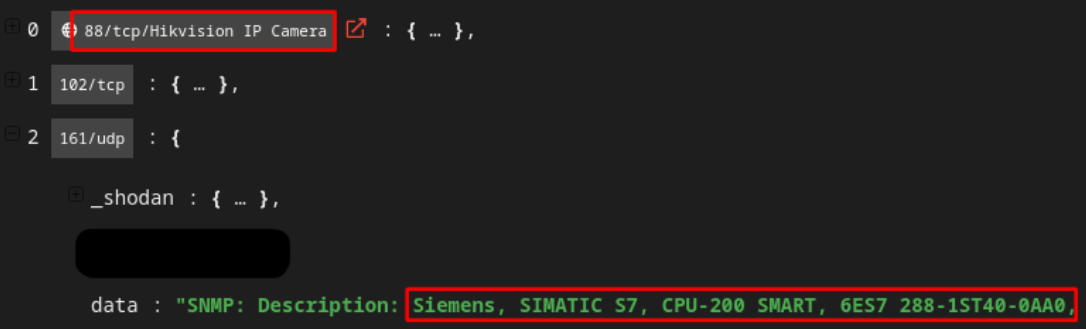}
    \caption{An example of an irregular port combination, presenting as both Siemens device and Hikvision IP camera.}
    \label{fig:irregular_combo}
\end{figure}

A range of positive honeypot conclusions can be attributed to irregular port configurations. Most were yielded through the `6ES7' search string and present convincing S7-info configurations, but further analyses reveals a host running multiple simultaneous honeypots. This is visible in Figure~\ref{fig:irregular_combo} where we can observe an address presenting both a Siemens device through SNMP on port 161 and a Hikvision IP camera on port 88. 

Many honeypots triggered the Honeyscore system by being hosted on the cloud. Since ICS honeypots commonly run on Linux cloud hosts the Local algorithm was also able to gauge them as honeypots. The distribution of hosts is visible in Figure~\ref{tab:cloud_providers}. 

\begin{table}
    \centering
    \resizebox{\linewidth}{!}{
    \begin{tabular}{c c c}
         Provider & No. of Occurrences & \% of Consensus  \\
         \midrule
         Alibaba Cloud & 2 & 0.06\%  \\
         Amazon AWS & 50 & 1.57\% \\
         DigitalOcean & 85 & 2.68\% \\
         Google Cloud & 11 & 0.35\% \\
         Linode Cloud & 18 & 0.57\% \\
         Microsoft Azure & 15 & 0.47\% \\
         Tencent Cloud & 4 & 0.12\% \\
         Vultr & 6 & 0.19\% \\
         \bottomrule
    \end{tabular}
    }
    \caption{Distribution of cloud providers across the consensus results with number of occurrences and percentage of the results occupied.}
    \label{tab:cloud_providers}
\end{table}

\subsection{Contention Results}

\begin{figure}
    \centering
    \resizebox{\linewidth}{!}{
    \includegraphics[scale=0.39]{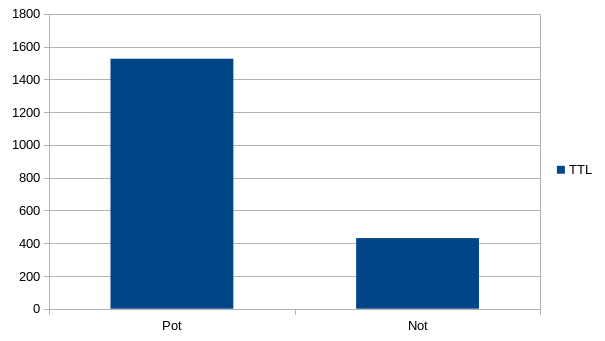}}
    \caption{Distribution between entries the TTL algorithm deemed a honeypot or not.}
    \label{fig:contention_pot_or_not}
\end{figure}

Contention results present a number of inconsistencies in both TTL and Shodan's conclusions. Figure~\ref{fig:contention_pot_or_not} displays the distribution between the devices the TTL algorithm concluded was a honeypot or a real device while Shodan concluded the opposite. In 431 cases, the TTL heuristic identified the device as a real while Shodan concluded that it was a honeypot. Manual analyses thend to lend credibility to Shodan's conclusion, with most entries displaying known honeypot strings or cloud host providers (Table~\ref{tab:contention_clouds} despite reconstructed TTLs falling within the 0 to 60 or 200 to 250 range. 

\begin{table}
    \centering
    \resizebox{\linewidth}{!}{
    \begin{tabular}{c c c}
         Host & No. of Occurrences & \% of Contention  \\
         \midrule
         Alibaba Cloud & 1 & 0.05\% \\
         Amazon AWS & 76 & 3.88\% \\
         DigitalOcean & 90 & 4.60\% \\
         Google Cloud & 34 & 1.73\% \\
         Hanauer Landstraße & 1 & 0.05\% \\
         Linode & 24 & 1.22\% \\
         Microsoft Azure & 11 & 0.56\% \\
         Oracle & 1 & 0.05\% \\
         Constant & 1 & 0.05\% \\
         Vultr & 3 & 0.15\% \\
         \bottomrule
    \end{tabular}
    }
    \caption{Occurrences of cloud hosts in the contention dataset.}
    \label{tab:contention_clouds}
\end{table}

Of the 1526 contentious results, the TTL algorithm deems 64 to be a honeypot which Shodan does not identify, yet which present clear known honeypot strings when analysed. These results are only marked with cloud tags while displaying the `Technodrome' and `Mouser Factory' strings. The remaining 1472 addresses are all devices marked as legitimate ICS devices by Shodan, 629 of which contained hardware strings for which we had previously recorded the TTL value. Manual analysis of these 629 entries reveals little in the way of identifiable honeypot strings with only some results presenting questionable host organisations in the form of mobile internet providers. Since different countries often possess their own unique internet service providers an unknown provider cannot constitute an immediately revealing characteristic.

The remaining 842 results are devices that exhibit untested hardware strings. Among these are configurations that display inconsistent hardware and modules (implying the use of input/output modules) and differing manufacturers utilising Siemens CPUs in their own devices. Relevant hardware strings of these devices are recorded in Table~\ref{tab:hardware_strings}. The ET200-S and its variants are seen numerous times paired with other devices, notably S7-300s and INSEVIS devices. These pairings alongside the unstudied progression of TTL values in Siemens devices makes it difficult to determine the exact cause of the contention. 

\begin{table}
    \centering
    \resizebox{\linewidth}{!}{
    \begin{tabular}{c c c c}
         Manufacturer & Range & Model & Average TTL   \\
         \midrule
         INSEVIS & PC35xV & 315-2EH14-0AB0 & 64.33 \\
         \hline
         VIPA & 300S & 315-4NE12-0110 & 65.00\\
         \hline
         \multirow{24}{*}{Siemens} & \multirow{1}{*}{ET200-S} & 151-8AB01-0AB0 & 69.48 \\ 
         & \multirow{10}{*}{S7-300} & 313-5BF03-0AB0 & 71.74\\ 
         & & 313-5BG04-0AB0 & 69.05 \\ 
         & & 314-1AG13-0AB0 & 69.84 \\
         & & 314-6EH04-0AB0 & 69.52 \\
         & & 315-2EH13-0AB0 & 63.56 \\
         & & 315-2AH14-0AB0 & 69.41 \\
         & & 315-2EH14-0AB0 & 68.14 \\
         & & 315-2FJ14-0AB0 & 68.10 \\
         & & 318-2AJ00-0AB0 & 70.51 \\
         & \multirow{14}{*}{S7-1200} & 211-1AE40-0XB0 & 69.80 \\ 
         & & 211-1HE40-0XB0 & 66.05 \\
         & & 212-1AD30-0XB0 & 64.61 \\ 
         & & 212-1AE40-0XB0 & 70.35 \\
         & & 212-1HE40-0XB0 & 69.22 \\
         & & 214-1AE30-0XB0 & 69.60 \\
         & & 214-1AF40-0XB0 & 73.07 \\
         & & 214-1AG31-0XB0 & 71.09 \\
         & & 214-1BG40-0XB0 & 71.05 \\
         & & 214-1HE30-0XB0 & 82.70 \\
         & & 214-1HG31-0XB0 & 87.12 \\
         & & 214-1HG40-0XB0 & 81.75 \\
         & & 215-1BG40-0XB0 & 89.04 \\
         & & 215-1HG40-0XB0 & 82.06 \\
         & & 217-1AG40-0XB0 & 71.25 \\
         & & \\
    \end{tabular}
    }
    \caption{Recorded hardware strings with unknown TTL default values alongside the averages from the contention dataset.}
    \label{tab:hardware_strings}
\end{table}

Analysis of each unknown device strings average TTL creates arguments in the Local conclusions favour. Of the 26 unknowns only a singular device presents below the Linux default TTL of 64 by a meagre .44. Further questions are raised when considering this model appears several times in the consensus dataset, being marked each time as not a honeypot by both algorithms. There are two hypotheses to explain such occurrences, the first being that Shodan is wrong and its algorithm has failed to identify multiple well-configured honeypots. The second is that these devices present an atypical TTL unlike those we have recorded.

\section{Discussion} \label{sec:discuss}

The potential of the TTL value as a honeypot identifying heuristic is displayed by the results discussed in Section~\ref{sec:results}. This is not to say it is a perfect method. This heuristic would see better performance when integrated into a tool utilising multiple weighted heuristics. This will require additional research concerning TTL values in Siemens devices, exploring the variations across generations, models, and manufacturers. Furthermore, the interaction between the CPU and its modules needs to be assessed. A Siemens device often comes with a Siemens AG network interface, but the TTL value could be changed by an additional I/O module. 

Despite this, an algorithm using only the TTL heuristic was demonstrated successfully identifying both honeypots and real devices 3176 times without access to additional characteristics. This is notable considering the algorithm only had to interact with a device twice while generating an extremely limited amount of traffic. This allows for attackers to reveal less of their capabilities and potentially avoid triggering the honeypot all-together. All identification was done agnostic of easily revealing characteristics only retrievable through scanning. 

The TTL algorithm was also able to identify 64 honeypots that Shodan had not, verified through the presence of known honeypot strings. The reason for this is not entirely known. It is not unreasonable to assume that the Shodan honeyscore system is imperfect but it must be questioned why characteristics flagged in the consensus data set were not flagged in these. The prevailing hypothesis is that the honeyscoring system operates a set amount of time after the ingestion of the scan. This would mean a period of time exists where honeypots are not yet marked as such. 

Additional issues with Shodan arise when using its API to fetch search string results. When using multiple search vectors it is very easy to use up the search credits quickly. This paired with the unreliability of download stream integrity makes for a difficult process. Furthermore, the JSON entries yielded by the search strings only include the port with the meta-data that contained a match for the search string. This means in order to further analyse port profiles additional queries must be made. It is possible to do this manually and not use any API credits, but when working with hundreds of results this becomes unfeasible. If this study is to be replicated continuously using other heuristics then alternative platforms or an appropriate licensing scheme should be utilised. Furthermore, Shodans API only provides its results in a JSON GZIP format that deviates from the valid json format, making it only easily parsed by the Shodan API command-line tools limited parsing functions. In order to perform more detailed analyses in future studies, the results from the queries should be parsed out into a more valid and workable format using a more sophisticate algorithm for ingestion. 

After generating and processing results, necessary improvements were acknowledged for the Ping and TraceRoute functions. Creating a function that is capable of performing a TraceRoute and capturing the final reply packet would consolidate the two functions into a singular process, therefore reducing run-time and potentially allowing for the easier implementation of multi-threaded processing. Similarly, the use of the TTL heuristic should be optimised to allow for a higher maximum TTL and reduce the plateau issue identified in the consensus results. 

The issue of irregular TTL is easily addressed on the typical Linux platform. TTL can be changed for a single session using the command:

\begin{lstlisting}[breaklines=true]
$ sysctl -w net.ipv4.ip_default_ttl=(VALUE)
\end{lstlisting}

This is easily integrated into scripts for starting device configurations. The not so trivial part is determining which TTL value is necessary for the device being emulated, but it can be assumed that so long as a device isn't immediately identified by the Linux default of 64 that anything can be used to replace it. Given that S7 devices seem to have TTL values atypical to conventional network devices, it can also be assumed that acceptable values would be 30, 60, and 255. These would move the device anyway from definite identification and improve its deception. In the event that a more consistent solution is necessary, like a long term honeypot that needs to retain configuration across restarts, the following variable can be set in the `/etc/sysctl.conf' file:

\begin{lstlisting}
net.ipv4.ip_default_ttl=(VALUE)
\end{lstlisting}

When considering the outcomes of this research, further research is needed to develop the deception of industrial honeypots. This need not concern solely complex features like protocols and physics simulations but also minor features like network behaviours and pseudo-side-channel methods like identifying operating systems trough exposed ports. These minor configurations problems can be overlooked due to their simplicity and require a greater focus in literature. This complacency could give sophisticated attackers an easier time in their identification of devices.  

\section{Conclusion} \label{sec:conclusion}
This study demonstrated the use of IP frame time-to-live values for identifying industrial control system honeypots. Shodan was leveraged for retrieving lists of IP addresses that correlated with common Siemens device features and known honeypot strings. Analysis revealed that real devices can be correctly distinguished from honeypots using TTL value alone and as a heuristic. Future work will explore the development of a scanning tool that makes use of TTL values alongside other revealing honeypot characteristics as weighted heuristics. This will enable results to determine the likelihood a device is a honeypot rather than relying on a strict dichotomy. Further work can also include the implementation of a TTL setting function in known mainstream honeypot implementations like ConPot in order to develop interactivity and cover a small but easy method of identification. 

\section{Acknowledgements}
Funding for this study was provided by the UKRI.
\bibliographystyle{abbrv}
\bibliography{main.bib}
\end{document}